\documentclass[twocolumn,floatfix,prb]{revtex4}
\usepackage{graphicx}
\usepackage{amsmath}
\begin{document}
\title{Quantum teleportation by particle-hole annihilation in the Fermi sea}
\author{C. W. J. Beenakker and M. Kindermann}
\affiliation{Instituut-Lorentz, Universiteit Leiden, P.O. Box 9506, 2300 RA
Leiden, The Netherlands}
\date{4 July 2003}
\begin{abstract}
A tunnel barrier in a degenerate electron gas was recently discovered as a source of entangled particle-hole excitations. The entanglement is produced by elastic tunneling events, without requiring electron-electron interactions. Here we investigate the inverse process, the annihilation of an electron and a hole by elastic scattering. We find that this process leads to {\em teleportation\/} of the (unknown) state of the annihilated electron to a second, distant electron --- if the latter was previously entangled with the annihilated hole. We propose an experiment, involving low-frequency noise measurements on a two-dimensional electron gas in a high magnetic field, to detect teleportation of electrons and holes in the two lowest Landau levels.
\end{abstract}
\pacs{03.65.Ud, 03.67.Mn, 73.43.-f, 73.50.Td}
\maketitle

Teleportation is the disembodied transport of a quantum mechanical state between two locations that are only coupled by classical (incoherent) communication \cite{Ben93}. What is required is that the two locations share a previously entangled state. Teleportation has the remarkable feature that the teleported state need not be known. It could even be undefined as a single-particle state, which happens if the teleported particle is entangled with another particle that stays behind. Teleportation then leads to ``entanglement swapping'' \cite{Yur92,Zuk93}: Pre-existing entanglement is exchanged for entanglement between two parties that have never met.

Experiments with photons \cite{Bou97,Pan98} have demonstrated that teleportation can be realized in practice. Only {\em linear\/} optical elements are needed, if one is satisfied with a success probability less than unity \cite{Vai99,Lut99}. Such non-deterministic teleportation plays an essential role in proposals for a quantum computer based entirely on linear optics \cite{Kni01}. A central requirement for nontrivial logical operations is that the linear elements (beam splitters, phase shifters) are supplemented by single-photon sources and single-photon detectors, which effectively introduce nonlinearities. 

Teleportation of electrons has not yet been realized. The analogue of teleportation by linear optics would be teleportation of free electrons, that is to say, teleportation using only single-particle Hamiltonians. Is that possible? A direct translation of existing linear optics protocols would require single-electron sources and single-electron detectors. Such devices exist \cite{Hol02,Wie03}, but not for free electrons --- they are all based on the Coulomb interaction in quantum dots. In this article we would like to propose an alternative.

\begin{figure}
\includegraphics[width=8cm]{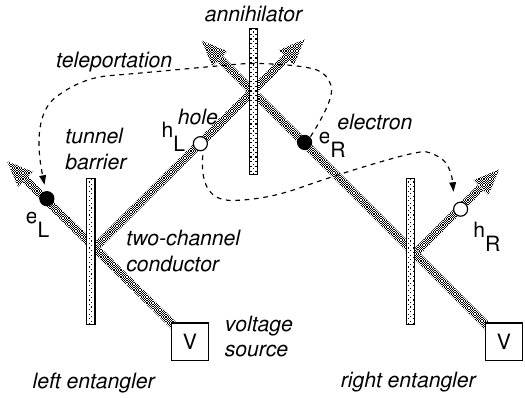}
\caption{
Schematic description of teleportation by particle-hole annihilation. A voltage $V$ applied over a tunnel barrier produces pairs of entangled electron-hole pairs in the Fermi sea. One such pair $(e_{L},h_{L})$ is shown at the left. For a simplified description we assume spin entanglement in the state $(|\!\uparrow\uparrow\rangle+|\!\downarrow\downarrow\rangle)/\sqrt{2}$, where the first arrow refers to the electron spin and the second arrow to the hole spin. (The more general situation is analyzed in the text.) A second electron $e_{R}$ is in an unknown state $\alpha|\!\uparrow\rangle+\beta|\!\downarrow\rangle$. The electron $e_{R}$ can annihilate with the hole $h_{L}$ by tunneling through the barrier at the center. If it happens, and is detected, then the state of $e_{L}$ collapses to the state of $e_{R}$. (Notice that $|\!\uparrow\rangle$ annihilates with $|\!\uparrow\rangle$ and $|\!\downarrow\rangle$ annihilates with $|\!\downarrow\rangle$, so $e_{L}$ inherits the coefficients $\alpha$ and $\beta$ of $e_{R}$ after its annihilation.) The diagram shows a second entangler at the right, to perform two-way teleportation (from $e_{R}$ to $e_{L}$ and from $h_{L}$ to $h_{R}$).  This leads to entanglement swapping: $e_{L}$ and $h_{R}$ become entangled after the annihilation of $h_{L}$ and $e_{R}$. 
\label{teleport_schematic}
}
\end{figure}

The key observation is that the annihilation of a particle-hole pair in the Fermi sea {\em teleports\/} these quasiparticles to a distant location, if entanglement was established beforehand. This two-way teleportation scheme is explained in Fig.\ \ref{teleport_schematic}. The two entanglers are taken from Ref.\ \onlinecite{Bee03}. There it was shown that the ``no-go'' theorem for entanglement production by linear optics does not carry over to electrons. In linear optics no entanglement can be generated from sources in thermal equilibrium \cite{Kim02,Xia02}. For electrons, on the contrary, this is possible. A tunnel barrier in a two-channel conductor creates entangled electron-hole pairs in the Fermi sea, using only single-particle elastic scattering. No single-electron sources are needed. Our proposal for teleportation uses the inverse process, the annihilation of a particle-hole excitation by elastic scattering.

\section*{The simplest case}

The analysis is simplest for the entangled state
\[
(|\!\uparrow\rangle_{e}|\!\uparrow\rangle_{h}+ |\!\downarrow\rangle_{e}|\!\downarrow\rangle_{h})/\sqrt{2}.
\]
The subscripts $e$ and $h$ refer, respectively, to electron and hole at two distant locations. The particle to be teleported is another electron, in the state $\alpha|\!\uparrow\rangle_{e'}+\beta|\!\downarrow\rangle_{e'}$ (with $|\alpha|^{2}+|\beta|^{2}=1$). The second electron $e'$ may tunnel into the empty state representing the hole $h$, but only if the spins match. If $t$ denotes the tunneling amplitude, then this happens with probability $\frac{1}{2}|\alpha|^{2}|t|^{2}+\frac{1}{2}|\beta|^{2}|t|^{2}=\frac{1}{2}|t|^{2}\ll 1$. The resulting annihilation of the two quasiparticle excitations collapses the combined state
\[
(\alpha|\!\uparrow\rangle_{e'}+\beta|\!\downarrow\rangle_{e'})
(|\!\uparrow\rangle_{e}|\!\uparrow\rangle_{h}+ |\!\downarrow\rangle_{e}|\!\downarrow\rangle_{h})/\sqrt{2}
\]
to the state $\alpha|\!\uparrow\rangle_{e}+\beta|\!\downarrow\rangle_{e}$, so the state of the second electron $e'$ is teleported to the first electron $e$ at a distant location.

The usual limitations \cite{Ben93} of teleportation apply. Since tunneling is an unpredictable stochastic event, it has to be detected and communicated (by classical means) to the distant location. There is therefore no instantaneous transfer of information. And since the electron has to be annihilated in order to be teleported, its (unknown) state can not be copied. Teleportation by particle-hole annihilation thus presents a rather dramatic demonstration of the no-cloning theorem of quantum mechanics \cite{Woo82,Die82} in action.

A major obstacle to teleportation in the solid state is the requirement of fast time-resolved detection. To circumvent this difficulty we identify a low-frequency noise correlator that demonstrates the entanglement swapping resulting from two-way teleportation. Two-way teleportation means that upon annihilation the electron and hole are teleported to opposite ends of the system. The noise correlator measures the degree of entanglement at the two ends. This demonstrates teleportation if the two ends are not connected by any phase-coherent path. An additional benefit from this detection scheme is that the degree of entanglement is invariant under local unitary transformations, so no precise knowledge and control of the phase relation of the entangled electrons and holes is required.

\section*{The general case}

\begin{figure*}
\includegraphics[width=16cm]{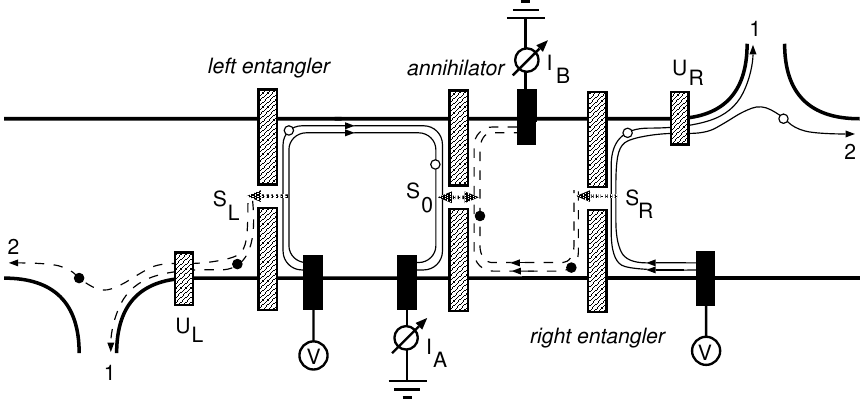}
\caption{
Proposed realization of the teleportation scheme of Fig.\ \protect\ref{teleport_schematic}, using edge channels in the quantum Hall effect. The thick black lines indicate the boundaries of a two-dimensional electron gas, connected by Ohmic contacts (black rectangles) to a voltage source $V$ or to ground. A strong perpendicular magnetic field ensures that the transport in an energy range $eV$ above the Fermi level takes place in two edge channels, extended along a pair of equipotentials (thin solid and dashed lines, with arrows that give the direction of propagation). These edge channels realize the two-channel conductors of Fig.\ \protect\ref{teleport_schematic}, with the Landau level index $n=1,2$ playing the role of the spin index $\uparrow,\downarrow$. Solid lines signify predominantly filled edge channels with hole excitations (open circles), while dashed lines signify predominantly empty edge channels with particle excitations (black dots). The beam splitters of Fig.\ \protect\ref{teleport_schematic} are formed by split gate electrodes (shaded rectangles), through which the edge channels may tunnel (dashed arrows, scattering matrices $S_{L},S_{R},S_{0}$). The annihilation of the particle-hole excitation at the central beam splitter is detected through the currents $I_{A}$ and $I_{B}$. Entanglement swapping resulting from two-way teleportation is detected by the violation of a Bell inequality. This requires two gate electrodes to locally mix the edge channels (scattering matrices $U_{L}$, $U_{R}$) and two pair of contacts $1,2$ to separately measure the current in each transmitted and reflected edge channel. Notice that there are no paths connecting the left and right ends of the conductor, so a demonstration of entanglement between the two ends is indeed a demonstration of teleportation.
\label{teleport_QHE}
}
\end{figure*}

We now proceed to the general formulation of teleportation by particle-hole annihilation. We follow Ref.\ \onlinecite{Bee03} by focusing on a particular implementation using edge channels in the quantum Hall effect regime (see Fig.\ \ref{teleport_QHE}). The entangled degree of freedom is the Landau level index $n=1,2$, which labels the two occupied edge channels near the Fermi energy $E_{F}$. Electrons are incident in a narrow range $eV$ above $E_{F}$ from two voltage sources. We write the incoming state
\begin{equation}
|\Psi_{\rm in}\rangle=a^{\dagger}_{L,1}a^{\dagger}_{L,2}a^{\dagger}_{R,1} a^{\dagger}_{R,2}|0\rangle \label{Psiindef}
\end{equation}
in second quantized form, in terms of operators $a_{L,n}^{\dagger}$ ($a_{R,n}^{\dagger}$) that excite the $n$-th edge channel at the left (right) voltage source. (The excitation energy $0<\varepsilon<eV$ is omitted for simplicity.) The vacuum state $|0\rangle$ represents the Fermi sea at zero temperature (all states below $E_{F}$ occupied, all states above $E_{F}$ empty).

Scattering matrices $S_{L}$, $S_{R}$ (for the left and right barriers acting as entanglers) and $S_{0}$ (for the central barrier acting as annihilator) transform the incoming state $|\Psi_{\rm in}\rangle$ to the outgoing state $|\Psi_{\rm out}\rangle$. The full expression for $|\Psi_{\rm out}\rangle$ is lengthy, but we only need the terms that correspond to the annihilation of electron and hole at the central barrier. If electron and hole have annihilated, this implies that there are two filled edge channels in contact $A$ and two empty edge channels in contact $B$. So these terms can be extracted by the projection operator
\begin{equation}
{\cal P}=n_{A,1}n_{A,2}(1-n_{B,1})(1-n_{B,2}). \label{Pdef}
\end{equation}
We have introduced the number operator $n_{X,n}=b^{\dagger}_{X,n}b^{\vphantom{\dagger}}_{X,n}$, with $b^{\dagger}_{X,n}$ the creation operator for the $n$-th edge channel approaching contact $X=A,B$ in Fig.\ \ref{teleport_QHE}. 

The projected outgoing state,
\begin{eqnarray}
{\cal P}|\Psi_{\rm out}\rangle&=&\bigl(\alpha b^{\dagger}_{R,1}b^{\dagger}_{R,2}+\beta b^{\dagger}_{L,1}b^{\dagger}_{L,2}\nonumber\\
&&\mbox{}+\sum_{n,m=1,2}\gamma^{\vphantom{\dagger}}_{nm} b^{\dagger}_{L,n} b^{\dagger}_{R,m}\bigr)b^{\dagger}_{A,1}b^{\dagger}_{A,2}|0\rangle, \label{PPsiout}
\end{eqnarray}
contains three types of contributions: 1. A term $\propto\alpha$ describing two filled edge channels to the right of the right barrier (creation operator $b^{\dagger}_{R,n}$); 2. A term $\propto\beta$ describing two filled edge channels to the left of the left barrier (creation operator $b^{\dagger}_{L,n}$); 3. A sum of four terms $\propto\gamma_{nm}$ describing one filled edge channel at the left and one at the right. The coefficients $\alpha,\beta,\gamma_{nm}$ are given in terms of the 
reflection and transmission matrices of the three barriers:
\begin{eqnarray}
&&\alpha=(r_{R}\sigma_{y}r_{R}^{T})_{12}(r_{0}r_{L} \sigma_{y}r_{L}^{T}r_{0}^{T})_{12},\label{alphadef}\\
&&\beta=(t_{L}\sigma_{y}t_{L}^{T})_{12}(t_{0}t_{R} \sigma_{y}t_{R}^{T}t_{0}^{T})_{12},\label{betadef}\\
&&\gamma=t_{L}\sigma_{y} r_{L}^{T}r_{0}^{T}\sigma_{y} t_{0}t_{R}\sigma_{y} r_{R}^{T}. \label{gammadef}
\end{eqnarray}
The superscript $T$ indicates the transpose of a matrix and
\begin{equation}
\sigma_{y}=\left(\begin{array}{cc}
0&-i\\i&0
\end{array}\right)\label{sigmaydef}
\end{equation}
is a Pauli matrix. If we denote by $t\ll 1$ the order of magnitude of the tunneling amplitudes, then $\alpha={\cal O}(t^{0})$, $\beta={\cal O}(t^{6})$, and $\gamma={\cal O}(t^{3})$, so it is justified to neglect $\beta$ relative to $\gamma$.

To identify the entangled electron-hole excitations we transform from particle to hole operators at contact $A$ and to the right of the right barrier: $b^{\dagger}_{A,n}\rightarrow c_{A,n}$, $b^{\dagger}_{R,n}\rightarrow c_{R,n}$. The new vacuum state is $|0'\rangle=b^{\dagger}_{R,1}b^{\dagger}_{R,2}b^{\dagger}_{A,1}b^{\dagger}_{A,2}|0\rangle$. The projected outgoing state becomes, upon normalization,
\begin{eqnarray}
&&{\cal P}|\Psi_{\rm out}\rangle=\sqrt{w}|\Phi\rangle+\sqrt{1-w}|0'\rangle+{\cal O}(t^{6}),\label{wdef}\\
&&|\Phi\rangle=w^{-1/2}\sum_{n,m=1,2}(\gamma\sigma_{y})_{nm} b^{\dagger}_{L,n}c^{\dagger}_{R,m}|0'\rangle. \label{phidef}
\end{eqnarray}
It represents a superposition of the vacuum state and a particle-hole state $|\Phi\rangle$ with weight $w=\sum_{n,m}|\gamma_{nm}|^{2}$.

The degree of entanglement of $|\Phi\rangle$ is quantified by the concurrence \cite{Woo98}, which ranges from $0$ (no entanglement) to $1$ (maximal entanglement). The concurrence
\begin{equation}
{\cal C}=2\frac{\sqrt{\Gamma_{1}\Gamma_{2}}}{\Gamma_{1}+\Gamma_{2}} \label{Cdef}
\end{equation}
is determined by the eigenvalues $\Gamma_{1},\Gamma_{2}$ of the matrix product $\gamma\gamma^{\dagger}$. A simple expression for these two eigenvalues exists if the left and right barrier each have the same tunnel probability for the two edge channels: $T_{L,1}=T_{L,2}$, $T_{R,1}=T_{R,2}$, with $T_{X,n}$ an eigenvalue of $t^{\vphantom{\dagger}}_{X}t^{\dagger}_{X}$. In this symmetric case the left and right barrier each produce maximally entangled electron-hole pairs \cite{Bee03}. The concurrence (\ref{Cdef}) then depends only on the tunnel probabilities $T_{0,n}$ of the central barrier, ${\cal C}=2(T_{0,1}T_{0,2})^{1/2}(T_{0,1}+T_{0,2})^{-1}$. If the central tunnel barrier is also symmetric ($T_{0,1}=T_{0,2}$), then ${\cal C}=1$, so the electron at the far left and the hole at the far right are maximally entangled. The two-way teleportation following particle-hole annihilation has therefore led to full entanglement swapping.

\section*{How to detect it}

The entanglement swapping can be detected by correlating the current fluctuations $\delta I_{L,n}$ and $\delta I_{R,n}$ in the $n$-th edge channel at the left and right end of the system. The correlator $\langle\delta I_{L,n}\delta I_{R,m}\rangle$ is zero, because there is no direct path between the two ends. A nonzero value is obtained by correlating with the current fluctuations $\delta I_{X}=\delta I_{X,1}+\delta I_{X,2}$ at the central contacts $X=A,B$. The third order correlator $\langle\delta I_{L,n}\delta I_{R,m}\delta I_{X}\rangle$ is still zero. The first nonvanishing correlator is of fourth order, for example $\langle\delta I_{L,n}\delta I_{R,m}\delta I_{A}\delta I_{B}\rangle$. We subtract the products of second order correlators to obtain the irreducible (cumulant) correlator at low frequencies,
\begin{equation}
\langle\!\langle\delta I_{L,n}(\omega_{1})\delta I_{R,m}(\omega_{2}) \delta I_{A}(\omega_{3})\delta I_{B}(\omega_{4})\rangle\!\rangle=2\pi\delta\bigl(\sum_{i=1}^{4} \omega_{i}\bigr)C_{nm}.\label{Cnmdef}
\end{equation}
It doesn't matter if $\delta I_{A}\delta I_{B}$ is replaced by $\delta I_{A}^{2}$ or $\delta I_{B}^{2}$, that only changes the correlator by a minus sign. Following Ref.\ \onlinecite{Lev93}, we have calculated $C_{nm}$ in terms of the transmission and reflection matrices, with the result
\begin{equation}
C_{nm}=2(e^{5}V/h)|M_{nm}|^{2},\;\;M=t^{\vphantom{\dagger}}_{L} r^{\dagger}_{L}r^{\dagger}_{0}t^{\vphantom{\dagger}}_{0} t^{\vphantom{\dagger}}_{R}r^{\dagger}_{R}.\label{Cnmresult}
\end{equation}

As in earlier work \cite{Cht02,Sam03}, we use low-frequency current correlators in the tunneling regime to detect entanglement through the violation of a Bell inequality \cite{Bel64}. We need the following rational function of correlators:
\begin{equation}
E=\frac{C_{11}+C_{22}-C_{12}-C_{21}}{C_{11}+C_{22}+C_{12}+C_{21}}. \label{Edef}
\end{equation}
By mixing the channels locally at the left and right end of the system, the transmission and reflection matrices are transformed as $t_{L}\rightarrow U_{L}t_{L}$, $r_{R}\rightarrow U_{R}r_{R}$, with unitary $2\times 2$ matrices $U_{L},U_{R}$. The Bell parameter \cite{Cla69}
\begin{equation}
{\cal E}= E(U_{L},U_{R})+E(U'_{L},U_{R})+E(U_{L},U'_{R})-E(U'_{L},U'_{R}) \label{CHSHdef}
\end{equation}
is maximized by a certain choice of $U_{L},U'_{L},U_{R},U'_{R}$ at the value \cite{Pop92}
\begin{equation}
{\cal E}_{\rm max}=2[1+4M_{1}M_{2}(M_{1}+M_{2})^{-2}]^{1/2}, \label{Emaxresult}
\end{equation} 
determined by the two eigenvalues $M_{1},M_{2}$ of the matrix product $MM^{\dagger}$.

To close the circle we need to show that $MM^{\dagger}$ and $\gamma\gamma^{\dagger}$ have the same eigenvalues, so that Eqs.\ (\ref{Cdef}) and (\ref{Emaxresult}) imply the one-to-one relation ${\cal E}_{\rm max}=2(1+{\cal C}^{2})^{1/2}$ between the concurrence and the maximal  value of the Bell parameter \cite{Gis91}. In general the two sets of eigenvalues $M_{1},M_{2}$ and $\Gamma_{1},\Gamma_{2}$ are different, but they become the same in the tunneling regime. Here is the proof.

In the tunneling regime the reflection matrices $r_{L},r_{R},r_{0}$ are close to being unitary. For any $2\times 2$ unitary matrix $U$ it holds that
\begin{equation}
\sigma_{y}U^{T}=e^{i\phi}U^{\dagger}\sigma_{y},\label{Uidentity}
\end{equation}
with $e^{i\phi}$ the determinant of $U$. With the help of this identity we may rewrite Eq.\ (\ref{gammadef}) as
\begin{equation}
\gamma=e^{i\phi}t^{\vphantom{\dagger}}_{L}r_{L}^{\dagger} r_{0}^{\dagger}t^{\vphantom{\dagger}}_{0} t^{\vphantom{\dagger}}_{R}r_{R}^{\dagger}\sigma_{y}= e^{i\phi}M\sigma_{y}.\label{gammaMrelation}
\end{equation}
Hence $\gamma\gamma^{\dagger}=MM^{\dagger}$, as we set out to prove.

A final remark: The Bell inequality states that $|{\cal E}|\leq 2$ for a local hidden-variable theory \cite{Bel64,Cla69}. We have not proven this statement for our fourth order correlator (although we do not doubt that it holds). What we have proven is that a measurement of the fourth order correlator can be used to determine the degree of entanglement, which is all we need for our purpose.

\section*{Discussion}

The invention of Bennett, Brassard, Cr\'{e}peau, Jozsa, Peres, and Wootters \cite{Ben93} teleports isolated and hence distinguishable particles, so it applies equally well to bosons (such as photons) as it does to fermions (such as electrons). However, the difficulty of isolating electrons in the solid state has so far prevented the realization of their ingenious idea. What we have shown here is that the existence of the Fermi sea makes it possible to implement teleportation of non-interacting fermions using sources in local thermal equilibrium --- something which is fundamentally forbidden for non-interacting bosons \cite{Kim02,Xia02}. Our fermions are not isolated electrons but particle-hole excitations created by tunneling events. The act of teleportation is the inverse process, the annihilation of the particle when it tunnels into the hole.

An advantage of working with particle-hole excitations in the Fermi sea is that no local control of single electrons is required. Indeed, the experiment proposed in Fig.\ \ref{teleport_QHE} does not need nanofabrication to isolate and manipulate electrons. A disadvantage is that the success rate of teleportation is small, because tunneling is a rare event. Since the particle-hole excitation survives if the tunneling attempt has failed, it should be possible to increase the teleportation rate by introducing more tunnel barriers in series. Such a more elaborate teleportation scheme might also be used to perform logical operations, along the lines of Ref.\ \onlinecite{Got99}.

\noindent{\bf Acknowledgements.}
This work was supported by the Dutch Science Foundation NWO/FOM and by the U.S. Army Research Office (Grant No. DAAD 19-02-0086).

\end{document}